\documentclass[nofootinbib, aps, prd, a4paper, 10pt, superscriptaddress, eqsecnum, showkeys]{revtex4-2}
\usepackage[a4paper, top=2cm, bottom=2cm, left=2.5cm, right=2.5cm]{geometry}

\usepackage{amsmath}
\usepackage{amsfonts}
\usepackage{amsthm}
\usepackage{bm}
\usepackage{mathrsfs}
\usepackage{braket}

\usepackage{graphicx}
\usepackage{booktabs}

\usepackage{xcolor}
\usepackage[pdfusetitle]{hyperref}
\hypersetup{colorlinks = true, allcolors = blue}

\usepackage{orcidlink}

\begin{document}

\title{ACT Data and Positive Running of the Spectral Index for Scalar Theory and Modified Gravity}
\author{S.D.~Odintsov\orcidlink{0000-0002-3529-7030}}
\email{odintsov@ieec.cat} \affiliation{ICREA, Passeig Luis
Companys, 23, 08010 Barcelona, Spain} \affiliation{Institute of
Space Sciences (IEEC-CSIC) C. Can Magrans s/n, 08193 Barcelona,
Spain}
\author{V.K.~Oikonomou\orcidlink{0000-0003-0125-4160}}
\email{voikonomou@gapps.auth.gr} \affiliation{Department of
Physics, Aristotle University of Thessaloniki, Thessaloniki 54124,
Greece} \affiliation{Center for Theoretical Physics, Khazar
University, 41 Mehseti Str., Baku, AZ-1096, Azerbaijan}

\begin{abstract}
In this work we address the possibility of having a positive
running of the spectral index in inflationary theories. The recent
ACT data indicate mildly that the running of the spectral index
might be positive, and several other physical indications point
out this possibility. If the running of the spectral index is
confirmed to be positive by future cosmic microwave background
experiments, this can rule out quite popular inflationary
scenarios. We investigate how it is possible to obtain a positive
running of the spectral index in the context of minimally coupled
scalar field gravity and modified gravity. For the modified
gravity we choose two mainstream and of string origin candidate
theories, $F(R)$ gravity and Einstein-Gauss-Bonnet gravity. In the
case of scalar field inflation and $F(R)$ gravity inflation, we
demonstrate the difficulties for obtaining a positive running of
the spectral index for a viable inflationary regime, so scalar
theories and $F(R)$ gravity are mostly compatible with the Planck
data. But nuanced scalar field scenarios can be compatible with
the ACT data and produce a positive running of the spectral index.
In the context of Einstein-Gauss-Bonnet theories which are
compatible with the GW170817 event, the running of the spectral
index can easily be positive while in parallel having a viable
inflationary era.
\end{abstract}

\maketitle

\section{Introduction}

Currently, the interest in theoretical physics is mainly connected
with the Cosmic Microwave Background (CMB) radiation related
experiments and with the future gravitational wave experiments.
The main focus is the fundamental idea of inflation
\cite{inflation2,inflation3,inflation4}, a unique theoretical
proposal that solves the observational problems of the standard
hot Big Bang theory. Currently no direct signal of inflation has
ever been confirmed, although the existence of nearly Gaussian
quantum fluctuations, that produce a nearly scale invariant power
spectrum, in combination with a experimentally confirmed flat
Universe, already confirms the quantum origin of the CMB
anisotropies. The trigger signal of an inflationary regime would
be the direct detection of the B-mode in the CMB, and the Simons
observatory \cite{SimonsObservatory:2019qwx} along with the
LiteBird experiment \cite{LiteBIRD:2022cnt} will directly aim
towards this. Another indirect confirmation of the inflationary
regime will be the detection of a stochastic primordial
gravitational wave background, and the future gravitational wave
experiments will operate in frequency bands that will probe the
inflationary gravitational waves
\cite{Hild:2010id,Baker:2019nia,Smith:2019wny,Crowder:2005nr,Smith:2016jqs,Seto:2001qf,Kawamura:2020pcg,Bull:2018lat,LISACosmologyWorkingGroup:2022jok}.
NANOGrav already verified a stochastic gravitational wave
background back in 2023 \cite{NANOGrav:2023gor}, but inflation
itself cannot generate such a signal
\cite{Vagnozzi:2023lwo,Oikonomou:2023qfz}. Thus the next ten years
will be extremely important for modern theoretical physics.

Recently, the Atacama Cosmology Telescope (ACT)
\cite{ACT:2025fju,ACT:2025tim} combined with the DESI data
\cite{DESI:2024uvr} stirred things up in CMB-related cosmology.
Already the DESI data found interesting phenomena
\cite{Lee:2025pzo,Ozulker:2025ehg,Kessler:2025kju,Nojiri:2025low,Vagnozzi:2019ezj},
that added up to the already known tensions of the
$\Lambda$-Cold-Dark-Matter model
\cite{Pedrotti:2024kpn,Jiang:2024xnu,Vagnozzi:2023nrq,Adil:2023exv,Bernui:2023byc,Gariazzo:2021qtg}.
But ACT offered another perspective that is in tension with the
Planck data \cite{Planck:2018jri}, and specifically the spectral
index was found to be,
\begin{equation}\label{act}
n_{\mathcal{S}}=0.9743 \pm 0.0034\, ,
\end{equation}
which is in $2\sigma$ tension with the Planck data. The
tensor-to-scalar ratio is constrained by the Planck and BICEP
experiments to be \cite{BICEP:2021xfz},
\begin{equation}\label{planck}
r<0.036
\end{equation}
at $95\%$ confidence. Although the $2\sigma$ is not a
statistically confident result, it is statistically non-ignorable.
Thus, it is somewhat vital to bare in mind that future experiments
may verify this ACT-Planck tension with higher confidence. Hence,
without forgetting that Planck-compatible inflationary models are
important, one must try to find models that can be compatible with
the ACT data. To this end, there exists a large literature that
try to reconcile standard and modified gravity inflation with the
ACT data, and for a stream of papers see Refs.
\cite{Kallosh:2025rni,Gao:2025onc,Liu:2025qca,Yogesh:2025wak,Yi:2025dms,Peng:2025bws,Yin:2025rrs,Byrnes:2025kit,
Wolf:2025ecy,Aoki:2025wld,Gao:2025viy,Zahoor:2025nuq,Ferreira:2025lrd,Mohammadi:2025gbu,Choudhury:2025vso,
Odintsov:2025wai,Q:2025ycf,Zhu:2025twm,Kouniatalis:2025orn,Hai:2025wvs,Dioguardi:2025vci,Yuennan:2025kde,
Kuralkar:2025zxr,Kuralkar:2025hoz,Modak:2025bjv,
Aoki:2025ywt,Ahghari:2025hfy,McDonough:2025lzo,Chakraborty:2025wqn,NooriGashti:2025gug,Yuennan:2025mlg,
Deb:2025gtk,Afshar:2025ndm,Ellis:2025zrf,Iacconi:2025odq,Yuennan:2025tyx,Wang:2025cpp,Qiu:2025uot,Wang:2025dbj,Asaka:2015vza,Oikonomou:2025htz,Choudhury:2025hnu,Singh:2025uyr,Kim:2025dyi,Peng:2026ofs,newref}
and references therein.

However, one interesting issue that is not mentioned in the
literature enough, perhaps due to its statistical insignificance
is the fact that the ACT data constrain the running of the
spectral index as follows \cite{ACT:2025tim},
\begin{equation}\label{runningactconstraint}
\frac{\mathrm{d}n_{\mathcal{S}}}{\mathrm{d}\ln k}=0.0062 \pm
0.0052\, ,
\end{equation}
so basically the central value is positive (actually both the
lower and maximum values are positive). This is in $1.2\sigma$
tension with the Planck data, however as it is pointed out in the
ACT release \cite{ACT:2025tim}, the constraints from the Planck
data prefer a slight negative running of the spectral index, while
the ACT and the combination P-ACT-LB prefer mildly a slight
positive running of the spectral index. In addition, when
combining the ACT with the Lyman-a forest and the Planck data, the
ACT release constraints disfavor a negative running of the
spectral index. Moreover, the recent literature also points out
that a positive running of the spectral index and the presence of
self-interacting dark radiation may remedy the Hubble tension at
$2.2\sigma$ level \cite{Garny:2026gcs}. Hence, we have mild, but
measurable indications that the running of the spectral index
might be positive. If this is confirmed at a higher statistical
confidence, many benchmark inflationary models would be
automatically ruled out. In this line of thinking, we shall
consider how a positive running may be achieved in the context of
modified gravity \cite{reviews1,reviews2,reviews3,reviews4} and in
standard minimally coupled scalar field inflation. For the
modified gravity we shall choose two mainstream theories of this
class of theories, namely the $F(R)$ gravity
\cite{Nojiri:2003ft,Capozziello:2005ku,Capozziello:2004vh,Capozziello:2018ddp,Hwang:2001pu,Bamba:2014mua,Nojiri:2006gh,Song:2006ej,Capozziello:2008qc,Bean:2006up,Capozziello:2012ie,Faulkner:2006ub,Olmo:2006eh,Sawicki:2007tf,Faraoni:2007yn,Carloni:2007yv,
Nojiri:2007as,Capozziello:2007ms,Deruelle:2007pt,Appleby:2008tv,Dunsby:2010wg,Oikonomou:2020oex,Oikonomou:2020qah}
and Einstein-Gauss-Bonnet theories
\cite{Hwang:2005hb,Cognola:2006sp,Nojiri:2005vv,Nojiri:2005jg,Satoh:2007gn,Yi:2018gse,Guo:2009uk,Guo:2010jr,Jiang:2013gza,vandeBruck:2017voa,Pozdeeva:2020apf,Vernov:2021hxo,Pozdeeva:2021iwc,Fomin:2020hfh,DeLaurentis:2015fea,Chervon:2019sey,Nozari:2017rta,Odintsov:2018zhw,Kawai:1998ab,Yi:2018dhl,vandeBruck:2016xvt,Maeda:2011zn,Ai:2020peo,Easther:1996yd,Codello:2015mba,Oikonomou:2021kql,Oikonomou:2022xoq,Odintsov:2020sqy,Oikonomou:2024etl,Fier:2025huc,Pozdeeva:2024ihc,Odintsov:2026cxz}.
Both theories are string motivated theories since, the one-loop
quantum corrected action of a single scalar field in its minimal
configuration (minimally or conformally coupled) contains the
following terms \cite{Codello:2015mba},
\begin{align}\label{quantumaction}
&\mathcal{S}_{eff}=\int
\mathrm{d}^4x\sqrt{-g}\Big{(}\Lambda_1+\Lambda_2
\mathcal{R}+\Lambda_3\mathcal{R}^2+\Lambda_4 \mathcal{R}_{\mu
\nu}\mathcal{R}^{\mu \nu}+\Lambda_5 \mathcal{R}_{\mu \nu \alpha
\theta}\mathcal{R}^{\mu \nu \alpha \theta}+\Lambda_6 \square
\mathcal{R}\\ \notag &
+\Lambda_7\mathcal{R}\square\mathcal{R}+\Lambda_8 \mathcal{R}_{\mu
\nu}\square \mathcal{R}^{\mu
\nu}+\Lambda_9\mathcal{R}^3+\mathcal{O}(\partial^8)+...\Big{)}\, ,
\end{align}
where the parameters $\Lambda_i$, $i=1,2,...,6$ are appropriate
dimensionful constants. As we show, having a positive running of
the spectral index, is a model-dependent issue, and generally
depends on the $F(R)$ gravity chosen and the scalar potential
chosen. In the case of the $F(R)$ gravity, we shall present a
general formulation to show how the running of the spectral index
can be obtained
\cite{Odintsov:2022rok,Odintsov:2020thl,Odintsov:2025eiv}, and how
its sign is a model-dependent issue that can be predicted only for
a handful of scenarios, due to lack of analyticity. Also for the
case of minimally coupled scalar field theory, we show that it
might be challenging to obtain a viable inflationary theory with a
positive running spectral index, in the context of standard
slow-roll inflation. On the contrary, the Einstein-Gauss-Bonnet
theories which are compatible with the GW170817 event, offer a
fruitful ground for viable inflationary phenomenology with a
positive running in the spectral index.

For this article, we shall assume a flat
Friedmann-Robertson-Walker spacetime, with metric,
\begin{equation}
\label{JGRG14} ds^2 = - dt^2 + a(t)^2 \sum_{i=1,2,3}
\left(dx^i\right)^2\, ,
\end{equation}
with $a(t)$ being the scale factor.

\section{Positive Running of the Spectral Index in Vacuum $F(R)$ Gravity: Problems and Model Building}

We start with the case of vacuum $F(R)$ gravity, and we shall
extract a general form for the running of the spectral index in
the Jordan frame. Note that in principle one can work out the
$F(R)$ gravity in the Einstein frame, but it is not always
possible to extract analytically the Einstein frame potential, as
in the case of power-law $F(R)$ gravity in the Jordan frame. The
vacuum $F(R)$ gravity in the Jordan frame has the following
action,
\begin{equation}\label{action1dse}
\mathcal{S}=\frac{1}{2\kappa^2}\int \mathrm{d}^4x\sqrt{-g}F(R),
\end{equation}
with $\kappa^2=8\pi G=\frac{1}{M_p^2}$ and $M_p$ denotes the
reduced Planck mass. By varying the action with respect to the
metric, we obtain the field equations,
\begin{equation}\label{eqnmotion}
F_R(R)R_{\mu \nu}(g)-\frac{1}{2}F(R)g_{\mu
\nu}-\nabla_{\mu}\nabla_{\nu}F_R(R)+g_{\mu \nu}\square F_R(R)=0\,
,
\end{equation}
where $F_R=\frac{\mathrm{d}F}{\mathrm{d}R}$. We can rewrite Eq.
(\ref{eqnmotion}) as follows,
\begin{align}\label{modifiedeinsteineqns}
R_{\mu \nu}-\frac{1}{2}Rg_{\mu
\nu}=\frac{\kappa^2}{F_R(R)}\Big{(}T_{\mu
\nu}+\frac{1}{\kappa^2}\Big{(}\frac{F(R)-RF_R(R)}{2}g_{\mu
\nu}+\nabla_{\mu}\nabla_{\nu}F_R(R)-g_{\mu \nu}\square
F_R(R)\Big{)}\Big{)}\, ,
\end{align}
and for the FRW metric (\ref{JGRG14}), the field equations become,
\begin{align}
\label{JGRG15} 0 =& -\frac{F(R)}{2} + 3\left(H^2 + \dot H\right)
F_R(R) - 18 \left( 4H^2 \dot H + H \ddot H\right) F_{RR}(R)\, ,\\
\label{Cr4b} 0 =& \frac{F(R)}{2} - \left(\dot H +
3H^2\right)F_R(R) + 6 \left( 8H^2 \dot H + 4 {\dot H}^2 + 6 H
\ddot H + \dddot H\right) F_{RR}(R) + 36\left( 4H\dot H + \ddot
H\right)^2 F_{RRR} \, ,
\end{align}
with $F_{RR}=\frac{\mathrm{d}^2F}{\mathrm{d}R^2}$, and
$F_{RRR}=\frac{\mathrm{d}^3F}{\mathrm{d}R^3}$. We shall assume
that the slow-roll conditions apply,
\begin{equation}\label{slowrollconditionshubble}
\ddot{H}\ll H\dot{H},\,\,\, \frac{\dot{H}}{H^2}\ll 1\, ,
\end{equation}
in which case the Ricci scalar becomes,
\begin{equation}\label{ricciscalarapprox}
R\sim 12 H^2\, .
\end{equation}
For the vacuum $F(R)$ gravity, the slow-roll indices are
\cite{Hwang:2005hb,reviews1},
\begin{equation}
\label{restofparametersfr}\epsilon_1=-\frac{\dot{H}}{H^2}, \quad
\epsilon_2=0\, ,\quad \epsilon_3= \frac{\dot{F}_R}{2HF_R}\, ,\quad
\epsilon_4=\frac{\ddot{F}_R}{H\dot{F}_R}\,
 ,
\end{equation}
and the observational indices  are \cite{reviews1,Hwang:2005hb},
\begin{equation}
\label{epsilonall} n_s=
1-\frac{4\epsilon_1-2\epsilon_3+2\epsilon_4}{1-\epsilon_1},\quad
r=48\frac{\epsilon_3^2}{(1+\epsilon_3)^2}\, .
\end{equation}
The tensor-to-scalar ratio for vacuum $F(R)$ gravity can be
simplified by using the Raychaudhuri equation,
\begin{equation}\label{approx1}
\epsilon_1=-\epsilon_3(1-\epsilon_4)\, ,
\end{equation}
and by employing the slow-roll conditions, we approximately have
$\epsilon_1\simeq -\epsilon_3$, so the spectral index takes the
form,
\begin{equation}
\label{spectralfinal} n_s\simeq 1-6\epsilon_1-2\epsilon_4\, ,
\end{equation}
and the tensor-to-scalar ratio takes the form $r\simeq 48
\epsilon_3^2$. So by combining the above with $\epsilon_1\simeq
-\epsilon_3$, we get,
\begin{equation}
\label{tensorfinal} r\simeq 48\epsilon_1^2\, .
\end{equation}
The slow-roll index $\epsilon_4$ is defined as,
\begin{equation}\label{epsilon41}
\epsilon_4=\frac{\ddot{F}_R}{H\dot{F}_R}=\frac{\frac{d}{d
t}\left(F_{RR}\dot{R}\right)}{HF_{RR}\dot{R}}=\frac{F_{RRR}\dot{R}^2+F_{RR}\frac{d
(\dot{R})}{d t}}{HF_{RR}\dot{R}}\, ,
\end{equation}
and by using the slow-roll assumptions, Eqs.  (\ref{rdot}) and
(\ref{epsilon41})  and also,
\begin{equation}\label{rdot}
\dot{R}\simeq -24H^3\epsilon_1\, ,
\end{equation}
we get,
\begin{equation}\label{epsilon4final}
\epsilon_4\simeq -\frac{24
F_{RRR}H^2}{F_{RR}}\epsilon_1-3\epsilon_1+\frac{\dot{\epsilon}_1}{H\epsilon_1}\,
,
\end{equation}
but $\dot{\epsilon}_1$ is,
\begin{equation}\label{epsilon1newfiles}
\dot{\epsilon}_1=-\frac{\ddot{H}H^2-2\dot{H}^2H}{H^4}=-\frac{\ddot{H}}{H^2}+\frac{2\dot{H}^2}{H^3}\simeq
2H \epsilon_1^2\, ,
\end{equation}
therefore $\epsilon_4$ gets the form,
\begin{equation}\label{finalapproxepsilon4}
\epsilon_4\simeq -\frac{24
F_{RRR}H^2}{F_{RR}}\epsilon_1-\epsilon_1\, .
\end{equation}
We introduce the dimensionless parameter $x$,
\begin{equation}\label{parameterx}
x=\frac{48 F_{RRR}H^2}{F_{RR}}\simeq \frac{4 F_{RRR}R}{F_{RR}}\, ,
\end{equation}
so the parameter $\epsilon_4$ becomes,
\begin{equation}\label{epsilon4finalnew}
\epsilon_4\simeq -\frac{x}{2}\epsilon_1-\epsilon_1\, .
\end{equation}
Substituting $\epsilon_4$ from Eq. (\ref{epsilon4finalnew}) in Eq.
(\ref{spectralfinal}), we get,
\begin{equation}\label{asxeto1}
n_s-1=-4\epsilon_1+x\epsilon_1\, ,
\end{equation}
Now we can derive a general expression for the running of the
spectral index in the case of vacuum $F(R)$ gravity. The
definition of the running of the spectral index is,
\begin{equation}\label{runningdef}
a_s=\frac{\mathrm{d} n_s}{\mathrm{d} \ln k}\, ,
\end{equation}
which can be rewritten as follows,
\begin{equation}\label{runningdef1}
a_s=\frac{\mathrm{d} n_s}{\mathrm{d} \ln k}=\frac{\mathrm{d}
n_s}{\mathrm{d}N}\frac{\mathrm{d}N}{\mathrm{d} \ln k}\, ,
\end{equation}
with $N$ being the $e$-foldings number. We use
$\frac{\mathrm{d}N}{\mathrm{d} \ln k}=\frac{1}{1-\epsilon_1}$, so
the running of the spectral index $a_s$ becomes,
\begin{equation}\label{runningdefmainfinal}
a_s=\frac{1}{1-\epsilon_1}\frac{\mathrm{d} n_s}{\mathrm{d}N}\, ,
\end{equation}
which is a general relation. We can further work out this
relation, so we get,
\begin{equation}\label{dndns}
\frac{\mathrm{d} n_s}{\mathrm{d}N}=\frac{\mathrm{d}
\epsilon_1}{\mathrm{d}N}\left( -4+x\right)+\frac{\mathrm{d}
x}{\mathrm{d}N} \epsilon_1\, ,
\end{equation}
but the term $\frac{\mathrm{d} \epsilon_1}{\mathrm{d}N}$ is equal
to,
\begin{equation}\label{depsilon1}
\frac{\mathrm{d}
\epsilon_1}{\mathrm{d}N}=\frac{\dot{\epsilon}_1}{H}\, ,
\end{equation}
and employing (\ref{epsilon1newfiles}) we get,
\begin{equation}\label{finalepsilon1dot}
\frac{\mathrm{d} \epsilon_1}{\mathrm{d}N}=2\epsilon_1^2\, .
\end{equation}
The term $\frac{\mathrm{d} x}{\mathrm{d}N}$ can be calculated
easily,
\begin{equation}\label{finalform}
\frac{\mathrm{d} x}{\mathrm{d}N}=-1152
\frac{F_{RRRR}}{F_{RR}}H^4-2x\epsilon_1+\frac{x^2}{2}\epsilon_1\,
,
\end{equation}
with $F_{RRRR}=\frac{\partial^4 F}{\partial R^4}$. Hence, we
finally have an analytic expression for $\frac{\mathrm{d}
n_s}{\mathrm{d}N}$, which basically controls the sign of the
running of the spectral index, and it is equal to,
\begin{equation}\label{runningFRgeneral}
\frac{\mathrm{d} n_s}{\mathrm{d}N}=2\epsilon_1^2\left(x-4
\right)-8
\frac{F_{RRRR}}{F_{RR}}R^2-2x\epsilon_1+\frac{x^2}{2}\epsilon_1 \,
.
\end{equation}
Now it is shown in Ref. \cite{Oikonomou:2025qub} that the de
Sitter scalaron mass of $F(R)$ gravity must be a monotonically
increasing function of the Ricci scalar or it must have an
extremum, during the slow-roll regime, where the curvature is
large. The de Sitter scalaron mass is equal to,
$$
m^2(R)=\frac{1}{3}\left(-R+\frac{F_R}{F_{RR}} \right)\, ,
$$
or in terms of the parameter $y$,
$$
m^2=\frac{R}{3}\left(-1+\frac{1}{y} \right)\, .
$$
where $y$ is defined as follows,
$$
y=\frac{R\,F_{RR}}{F_R}\, .
$$
Hence, the scalaron mass is,
$$
m^2(R)=\frac{1}{3}\left(-1+\frac{F_R }{F_{RR}R} \right)
$$
and it must be a monotonically increasing function of $R$, or to
have a local extremum, in order to have a small scalaron mass at
small curvatures,  and a large scalaron mass, at large curvatures,
which is motivated by the late-time behavior of the scalaron mass.
Hence, one must have,
$$
\frac{\partial m^2}{\partial R}\geq 0\, ,
$$
or in terms of $F(R)$,
$$
\frac{\partial m^2}{\partial
R}=-\frac{1}{12}\frac{F_R}{R\,F_{RR}}\,\frac{4\,R\,F_{RRR}}{F_{RR}}\geq
0 \, ,
$$
and finally in terms of $y$,
$$
\frac{\partial m^2}{\partial R}=-\frac{1}{12}\frac{x}{y}\geq 0 \,
.
$$
The scalaron mass must be positive or zero, hence we have,
$$
0< y \leq 1\, ,
$$
therefore in terms of $x$ and $y$ the above result to the
constraints,
$$
x\leq 0,\,\,\,0< y\leq 1\, .
$$
Also it was found in \cite{Oikonomou:2025qub} that $-1\leq x\leq
0$ in order for having a viable inflationary regime, therefore,
the function $x$ must be monotonically decreasing as a function of
the Ricci scalar $R$. Therefore, $x'(R)<0$, so by evaluating
$x'(R)$ we get,
\begin{equation}\label{xderivativer}
x'(R)=\frac{x}{R}-\frac{4 x^2}{R}+4\,R\,\frac{F_{RRRR}}{F_{RR}}\,
,
\end{equation}
thus in order to have $x'(R)<0$ one must have
$\frac{x}{R}>4\,R\,\frac{F_{RRRR}}{F_{RR}}$, but nothing
constraints $4\,R\,\frac{F_{RRRR}}{F_{RR}}$ which appears in Eq.
(\ref{runningFRgeneral}). For viable scenarios with $x\sim 0$, the
term $-8 \frac{F_{RRRR}}{F_{RR}}R^2$ appearing in
(\ref{runningFRgeneral}) crucially affects the sign of the running
of the spectral index. Thus it is difficult to extract a general
rule for the running of the spectral index for a general $F(R)$
gravity. This is possible for simple forms of $F(R)$ gravity, for
example $R^2$ gravity yields,
\begin{equation}\label{r2attractorsrunningofspectral}
a_s\simeq -\frac{8\epsilon_1^2}{1-\epsilon_1}\, .
\end{equation}
and the same can be done for simple $F(R)$ gravities. Thus for the
most known $F(R)$ gravities, the running of the spectral index is
negative. The whole analysis depends on the parameter $x$ and the
value of the term $-8 \frac{F_{RRRR}}{F_{RR}}R^2$ at first horizon
crossing. We have not found a viable Jordan frame example for
$F(R)$ gravity that can yield a positive running of the spectral
index.

\section{GW170817-compatible Einstein-Gauss-Bonnet Theories with Positive Running of the Spectral Index}

There exist two classes of EGB models developed in the literature
\cite{Oikonomou:2021kql,Oikonomou:2024etl} that can be compatible
with the GW170817 event and at the same time provide a viable
inflationary phenomenology. We shall consider here the models
developed in Ref. \cite{Oikonomou:2024etl}, which are known to be
compatible with the Planck data, and also compatible with the ACT
data \cite{Odintsov:2025wai}. As we now show, these models can
lead to a positive running of the spectral index, a feature that
is model-dependent though. Let us present in brief the formalism
of these models. The gravitational action of EGB theories is,
\begin{equation}
\label{action} \centering
S=\int{d^4x\sqrt{-g}\left(\frac{R}{2\kappa^2}-\frac{1}{2}\partial_{\mu}\phi\partial^{\mu}\phi-V(\phi)-\frac{1}{2}\xi(\phi)\mathcal{G}\right)}\,
,
\end{equation}
where $R$ is the Ricci scalar, $\kappa=\frac{1}{M_p}$ where $M_p$
is the reduced Planck mass, and $\mathcal{G}$ denotes the four
dimensional Gauss-Bonnet invariant,
$\mathcal{G}=R^2-4R_{\alpha\beta}R^{\alpha\beta}+R_{\alpha\beta\gamma\delta}R^{\alpha\beta\gamma\delta}$.
For the flat FRW spacetime, the variation of the action with
respect to both the scalar field and the metric, gives the
following field equations,
\begin{equation}
\label{motion1} \centering
\frac{3H^2}{\kappa^2}=\frac{1}{2}\dot\phi^2+V+12 \dot\xi H^3\, ,
\end{equation}
\begin{equation}
\label{motion2} \centering \frac{2\dot
H}{\kappa^2}=-\dot\phi^2+4\ddot\xi H^2+8\dot\xi H\dot H-4\dot\xi
H^3\, ,
\end{equation}
\begin{equation}
\label{motion3} \centering \ddot\phi+3H\dot\phi+V'+12 \xi'H^2(\dot
H+H^2)=0\, .
\end{equation}
We assume again a slow-roll era, which is realized by the
constraints,
\begin{equation}\label{slowrollhubble}
\dot{H}\ll H^2,\,\,\ \frac{\dot\phi^2}{2} \ll V,\,\,\,\ddot\phi\ll
3 H\dot\phi\, .
\end{equation}
The speed of the tensor perturbations for Einstein-Gauss-Bonnet
theories in a flat FRW spacetime is,
 \cite{Hwang:2005hb},
\begin{equation}
\label{GW} \centering c_T^2=1-\frac{Q_f}{2Q_t}\, ,
\end{equation}
where $Q_f$, $F$ and $Q_b$ stand for $Q_f=8 (\ddot\xi-H\dot\xi)$,
$Q_t=F+\frac{Q_b}{2}$, $F=\frac{1}{\kappa^2}$ and also $Q_b=-8
\dot\xi H$. The gravitational wave speed of the GW170817 event is,
\begin{align}
\label{GWp9} \left| c_T^2 - 1 \right| < 6 \times 10^{-15}\, .
\end{align}
so we can achieve this by assuming,
\begin{equation}\label{actualgw170817constraints}
\kappa^2\dot{\xi}H\ll 1,\,\,\,\kappa^2\ddot{\xi}\ll 1\, .
\end{equation}
In the formalism we consider, we further assume that,
$\kappa^2\dot{\xi}H^3\ll \kappa^2\dot{\phi}^2$ and
$\kappa^2\ddot{\xi}H^2\ll \kappa^2\dot{\phi}^2$, or wrapped up,
\begin{equation}\label{additionalconstraints}
\kappa^2\dot{\xi}H^3\ll
\kappa^2\dot{\phi}^2,\,\,\,\kappa^2\ddot{\xi}H^2\ll
\kappa^2\dot{\phi}^2\, .
\end{equation}
Thus the first field equation becomes,
\begin{equation}
\label{motion5} \centering H^2\simeq\frac{\kappa^2V}{3}\, ,
\end{equation}
and the Raychaudhuri equation takes the form,
\begin{equation}
\label{motion6} \centering \dot H\simeq-\frac{1}{2}\kappa^2
\dot\phi^2\, ,
\end{equation}
and the modified Klein-Gordon equation becomes,
\begin{equation}
\label{motion8} \centering \dot\phi\simeq
-\frac{12\xi'(\phi)H^4+V'}{3H}\, .
\end{equation}
The definition of the slow-roll indices for EGB inflation is the
following \cite{Hwang:2005hb},
\begin{align}
\centering \label{indices} \epsilon_1&=-\frac{\dot
H}{H^2}&\epsilon_2&=\frac{\ddot\phi}{H\dot\phi}&\epsilon_3&=0&\epsilon_4&=\frac{\dot
E}{2HE}&\epsilon_5&=\frac{Q_a}{2HQ_t}&\epsilon_6&=\frac{\dot
Q_t}{2HQ_t}\, ,
\end{align}
where $Q_a=-4\dot{\xi} H^2 $, $Q_b=-8\dot{\xi} H$,
$E=\frac{1}{(\kappa\dot\phi)^2}\left(
\dot\phi^2+\frac{3Q_a^2}{2Q_t}+Q_c\right)$, $Q_c=0$, $Q_d=0$,
$Q_e=-16 \dot{\xi} \dot{H}$, $Q_f=8\left(\ddot{\xi}-\dot{\xi}H
\right)$ and moreover $Q_t=\frac{1}{\kappa^2}+\frac{Q_b}{2}$. In
addition, the scalar spectral index and the tensor-to-scalar
ratio, in terms of the slow-roll indices are,
\begin{equation}\label{spectralindex}
n_{\mathcal{S}}=1+\frac{2 (-2
\epsilon_1-\epsilon_2-\epsilon_4)}{1-\epsilon_1}\, ,
\end{equation}
\begin{equation}\label{tensortoscalar}
r=\left |\frac{16 \left(c_A^3 \left(\epsilon_1-\frac{1}{4} \kappa
^2 \left(\frac{2
Q_c+Q_d}{H^2}-\frac{Q_e}{H}+Q_f\right)\right)\right)}{c_T^3
\left(\frac{\kappa ^2 Q_b}{2}+1\right)}\right |\, ,
\end{equation}
where $c_A$ denotes the sound speed of the scalar perturbations,
\begin{equation}\label{soundspeed}
c_A=\sqrt{\frac{\frac{Q_a Q_e}{\frac{2}{\kappa ^2}+Q_b}+Q_f
\left(\frac{Q_a}{\frac{2}{\kappa
^2}+Q_b}\right)^2+Q_d}{\dot{\phi}^2+\frac{3 Q_a^2}{\frac{2}{\kappa
^2}+Q_b}+Q_c}+1}\, .
\end{equation}
Also $e$-foldings number is defined as follows,
\begin{equation}
\label{efolds} \centering
N=\int_{t_i}^{t_f}{Hdt}=\int_{\phi_i}^{\phi_f}\frac{H}{\dot{\phi}}d\phi\,
,
\end{equation}
where $\phi_i$ and $\phi_f$ are the scalar field values at the
beginning and at the end of inflation. Due to Eqs. (\ref{efolds})
and (\ref{motion8}), we get,
\begin{equation}
\label{efolds1} \centering
N=\int_{\phi_i}^{\phi_f}\frac{2H^2}{12\xi'H^4+V'}d\phi\, ,
\end{equation}
and thus, due to Eq. (\ref{motion5}) we get,
\begin{equation}
\label{efoldsuncostrained} \centering
N=\int_{\phi_f}^{\phi_i}\frac{\kappa ^2 V(\phi )}{V'(\phi
)+\frac{4}{3} \kappa ^4 V(\phi )^2 \xi '(\phi )}d\phi\, .
\end{equation}
Moreover, a viable inflationary theory must have a viable
amplitude of scalar perturbations, compatible with Planck
\cite{Planck:2018jri} constraint
$\mathcal{P}_{\zeta}(k_*)=2.196^{+0.051}_{-0.06}\times 10^{-9}$.
For the case of Einstein-Gauss-Bonnet theories, the amplitude of
the scalar perturbations is defined as follows
\cite{Hwang:2005hb},
\begin{equation}\label{powerspectrumscalaramplitude}
\mathcal{P}_{\zeta}(k)=\left(\frac{k \left((-2
\epsilon_1-\epsilon_2-\epsilon_4) \left(0.57\, +\log \left(\left|k
\eta \right| \right)-2+\log (2)\right)-\epsilon_1+1\right)}{(2 \pi
) \left(z c_A^{\frac{4-n_{\mathcal{S}}}{2}}\right)}\right)^2\, ,
\end{equation}
where $z=\frac{a \dot{\phi} \sqrt{\frac{E(\phi )}{\frac{1}{\kappa
^2}}}}{H (\epsilon_5+1)}$ and
$\eta=-\frac{1}{aH}\frac{1}{-\epsilon_1+1}$, with everything
evaluated at the first horizon crossing. Using Eqs.
(\ref{motion5}), (\ref{motion6}) and (\ref{motion8}), we get,
\begin{equation}\label{epsilon1analytic}
\epsilon_1=\frac{4}{3} \kappa ^2 \xi '(\phi ) V'(\phi
)+\frac{V'(\phi )^2}{2 \kappa ^2 V(\phi )^2}+\frac{8}{9} \kappa ^6
V(\phi )^2 \xi '(\phi )^2\, .
\end{equation}
Analytical results can be obtained if we choose
\begin{equation}\label{couplingfunctionchoices3}
\xi'(\phi)=\frac{\lambda  V'(\phi )}{V(\phi )^2}\, ,
\end{equation}
and the speed of tensor perturbations takes the form,
\begin{equation}\label{gwspeedclass2}
c_T^2=\frac{-8 \lambda  (4 \lambda +3)^2 V(\phi ) V'(\phi )^2
V''(\phi )+10 \lambda  (4 \lambda +3)^2 V'(\phi )^4+27 V(\phi
)^4}{3 V(\phi )^2 \left(4 \lambda  (4 \lambda +3) V'(\phi )^2+9
V(\phi )^2\right)}\, ,
\end{equation}
and in addition, the first slow-roll index takes the form,
\begin{equation}\label{epsilon1formclass}
\epsilon_1=\frac{(4 \lambda +3)^2 V'(\phi )^2}{18 V(\phi )^2}\, ,
\end{equation}
while the $e$-foldings number takes the form,
\begin{equation}\label{finalinitialefoldings}
N=\int_{\phi_f}^{\phi_i} \frac{V(\phi )}{\frac{4}{3} \lambda
V'(\phi )+V'(\phi )} \mathrm{d}\phi\, .
\end{equation}
For the models of Eq. (\ref{couplingfunctionchoices3}), the
Gauss-Bonnet coupling grows large when the scalar potential
becomes smaller \cite{vandeBruck:2016xvt}. An interesting
ACT-compatible model with positive running of the spectral index
is the following,
\begin{equation}\label{viablepotentials3}
V(\phi)=\frac{M (\kappa  \phi )^2}{d+(\kappa  \phi )^2}\, ,
\end{equation}
which is known in the single scalar field literature as the radion
gauge inflation scalar potential (RGI) \cite{Fairbairn:2003yx}.
Using Planck units, we have,
\begin{equation}\label{xiphi}
\xi'(\phi)=\frac{2 d \lambda }{M \phi ^3}\, ,
\end{equation}
and the first slow-roll index becomes,
\begin{equation}\label{slowrollindexena}
\epsilon_1=\frac{2 d^2 (4 \lambda +3)^2}{9 \phi ^2 \left(d+\phi
^2\right)^2}\, ,
\end{equation}
and in addition, the $e$-foldings number takes the form,
\begin{equation}\label{integraln1}
N=\frac{\frac{3 d \phi ^2}{2}+\frac{3 \phi ^4}{4}}{8 d \lambda +6
d}\Big{|}_{\phi_f}^{\phi_i}\, .
\end{equation}
This model can yield a viable ACT-compatible phenomenology by
choosing $N=60$ by choosing the free parameters
$(d,\lambda,M)=(3,10^{-12},1.9474\times 10^{-10})$ in which case
the observational indices become, $n_{\mathcal{S}}=0.973871$, $r=
0.0112445$ and the predicted gravitational wave speed at first
horizon crossing is $|c_T^2-1|=9.12899\times 10^{-16}$. Finally,
the amplitude of the scalar perturbations becomes for this model
$\mathcal{P}_{\zeta}(k_*)=2.196\times 10^{-9}$. Now regarding the
running of the spectral index, which recall is defined as,
\begin{equation}\label{runningdef}
a_s=\frac{\mathrm{d} n_{\mathcal{S}}}{\mathrm{d} \ln k}\, ,
\end{equation}
we can rewrite $a_s$ as follows,
\begin{equation}\label{runningdef1}
a_s=\frac{\mathrm{d} n_{\mathcal{S}}}{\mathrm{d} \ln
k}=\frac{\mathrm{d}
n_{\mathcal{S}}}{\mathrm{d}N}\frac{\mathrm{d}N}{\mathrm{d} \ln
k}\, ,
\end{equation}
where $N$ is the $e$-foldings number. By using
$\frac{\mathrm{d}N}{\mathrm{d} \ln k}=\frac{1}{1-\epsilon_1}$, we
obtain a much more convenient for our needs expression for the
running of the spectral index $a_s$, which is finally written as,
\begin{equation}\label{runningdefmainfinal}
a_s=\frac{1}{1-\epsilon_1}\frac{\mathrm{d}
n_{\mathcal{S}}}{\mathrm{d}N}\, .
\end{equation}
Using this convenient formula, we can easily compute the running
of the spectral index for $N=60$ and
$(d,\lambda,M)=(3,10^{-12},1.9474\times 10^{-10})$, and it is
equal to $a_s=0.000443464$, which is positive and compatible with
the ACT constraint on the running of the spectral index
(\ref{runningactconstraint}). We found more models of this sort,
but we do not claim that the running is positive for all the
ACT-compatible models of this sort. This seems to be a
model-dependent feature.

\section{Minimally Coupled Scalar Field Theory and Positive Running of the Spectral Index}

Consider now a canonical minimally coupled single scalar field
inflation, with the action being,
\begin{equation}
S=\int d^4x \sqrt{-g} \left[ \frac{R}{2\kappa^2}
-\frac12(\partial\phi)^2 -V(\phi) \right].
\end{equation}
In slow-roll canonical single-field inflation, the scalar spectral
observables are conveniently expressed in terms of the Hubble
slow-roll parameters,
\begin{equation}
\epsilon_1 \equiv -\frac{\dot H}{H^2}, \qquad \epsilon_2 \equiv
\frac{d \ln \epsilon_1}{dN}, \qquad \epsilon_3 \equiv \frac{d \ln
\epsilon_2}{dN},
\end{equation}
where $N$ denotes as previously the $e$-foldings number. To first
order in the slow-roll expansion, the scalar spectral index and
its running are given by,
\begin{equation}
n_s - 1 = -2\epsilon_1 - \epsilon_2,
\end{equation}
and
\begin{equation}
\alpha_s \equiv \frac{dn_s}{d\ln k} \simeq -2\epsilon_1 \epsilon_2
- \epsilon_2 \epsilon_3.
\end{equation}
In order to have a positive running of the spectral index
therefore, one would need the following,
\begin{equation}
\epsilon_2 \left( 2\epsilon_1 + \epsilon_3 \right) < 0.
\end{equation}
Now, in many simple models, in order to obtain
\begin{equation}
\alpha_s > 0
\end{equation}
while maintaining a red-tilted power spectrum,
\begin{equation}
n_s < 1
\end{equation}
typically requires a negative $\epsilon_2$ parameter. Through the
relation,
\begin{equation}
n_s - 1 = -2\epsilon_1 - \epsilon_2,
\end{equation}
having a negative $\epsilon_2$, tends to force larger values of
$\epsilon_1$, which in turn imply a large tensor amplitude,
\begin{equation}
r = 16\epsilon_1.
\end{equation}
Consequently, simple analytic single-field realizations that
generate positive running, often predict values of $r$
incompatible with current Planck/BICEP constraint. This tendency
of the interplay between the parameters $\epsilon_2$ and
$\epsilon_1$, explains why viable positive-running scenarios
usually require more complicated dynamics, for example localized
features in the potential, or some inflection-point regions,
axion-like modulations, or violations of the slow-roll, or even
$k$-essence inflation. It might be possible though such mechanisms
to produce,
\begin{equation}
\alpha_s > 0
\end{equation}
while keeping the tensor-to-scalar ratio sufficiently suppressed.
Let us briefly consider a simple example of single scalar field
inflation, using a bottom-up approach, for example suppose that,
\begin{equation}
\epsilon_1(N) = \frac{A}{N^p},
\end{equation}
with a constant $A>0$ and $p>0$. One then obtains analytically,
\begin{equation}
\epsilon_2 = -\frac{p}{N},\,\,\, \epsilon_3 = -\frac{1}{N}.
\end{equation}
So the scalar spectral index becomes,
\begin{equation}
n_s - 1 = -\frac{2A}{N^p} + \frac{p}{N},
\end{equation}
while the running of the spectral index is,
\begin{equation}
\alpha_s = \frac{2Ap}{N^{p+1}} - \frac{p}{N^2}.
\end{equation}
The simplest case corresponds to $p=1$, for which,
\begin{equation}
\epsilon_1 = \frac{A}{N},
\end{equation}
and therefore we have,
\begin{equation}
n_s - 1 = -\frac{2A - 1}{N},
\end{equation}
and the running of the spectral index is,
\begin{equation}
\alpha_s = \frac{2A - 1}{N^2}.
\end{equation}
In this class of models, a positive running occurs naturally when,
\begin{equation}
2A - 1 > 0,
\end{equation}
thus imposing the observationally favored by ACT value,
\begin{equation}
n_s \simeq 0.974,
\end{equation}
and taking $N \simeq 60$, one finds,
\begin{equation}
A \simeq 1.28,
\end{equation}
which yields,
\begin{equation}
\alpha_s \simeq 4 \times 10^{-4}.
\end{equation}
However, the corresponding tensor-to-scalar ratio is,
\begin{equation}
r = 16\epsilon_1 = \frac{16A}{N},
\end{equation}
leading to,
\begin{equation}
r \simeq 0.34,
\end{equation}
which is observationally excluded. Thus, as in the case of simple
$F(R)$ gravity, the realization of a viable inflationary regime
with a positive running of the spectral index, can be cumbersome
and model-dependent. In general it is quite difficult to obtain
such a physical combination, so the model must be engineered for
achieving this combination of a viable inflationary regime with a
positive running of the spectral index. One sound example of this
sort in the context of minimally coupled scalar field theory is
the theory of singular analytic inflation, developed in Ref.
\cite{Oikonomou:2026mvp}. Since this article is devoted in viable
inflationary theories with positive running of their spectral
index, it is worth recalling the basic features of this singular
analytic inflation. The analytic solutions for the inflationary
regime are obtained if the kinetic energy of the scalar field
satisfies the following law,
\begin{equation}\label{mainequationpro}
\dot{\phi}^2=\frac{\beta}{\kappa^4}\left(\frac{H(t)}{H_0}\right)^{-m}\,
,
\end{equation}
with $\kappa^2=\frac{1}{M_p^2}$ and $M_p$ being the reduced Planck
mass, $\beta$ being a dimensionless parameter and also $H_0$
stands for some arbitrary mass scale, with its mass dimensions
being $[H_0]=[m]$. With $\gamma$,
\begin{equation}\label{gammadef}
\gamma=\frac{\beta}{\kappa^4}H_0^{m}\, ,
\end{equation}
the condition for the kinetic energy of the scalar field a FRW
spacetime reads,
\begin{equation}\label{mainequation}
\dot{\phi}^2=\gamma H(t)^{-m}\, .
\end{equation}
The assumption $\dot{\phi}^2 = \gamma H^{-m}$ is mainly motivated
by some attractor scaling solutions in the so-called k-mouflage
gravity, which is a class of modified-gravity theories,  with
several nonlinear derivative self-interactions which achieve
Vainshtein screening on the small scales, and they allow cosmic
acceleration during the late time of the Universe
\cite{Brax:2014wla}. For consistency, the assumptions on the
parameter $m$ are,
\begin{equation}\label{assumptionsonm}
m>2,\,\,\,m=2\ell,\,\,\,\mathrm{or}\,\,\,m=\frac{2\ell}{2
n_1+1},\,\,\ell, n_1=1,2,...\, .
\end{equation}
With (\ref{mainequation}), by solving the Raychaudhuri equation we
get the solution,
\begin{equation}\label{mainsolution}
H(t)=2^{-\frac{1}{m+1}} (m+1)^{\frac{1}{m+1}} \left(2
\omega-\gamma \kappa ^2 t\right)^{\frac{1}{m+1}}\, ,
\end{equation}
where $\omega$ is an integration constant with mass dimensions
$[\omega]=[m]^{m+1}$, which leads to the scale factor,
\begin{equation}\label{mainscalefactor}
a(t)=a_0 \exp \Big[-\frac{2^{-\frac{1}{m+1}} \,(m+1)^{n}
\,}{\gamma  \kappa ^2 (m+2)}\,\Big(2 \omega-\gamma \kappa ^2
t\Big)^{n}\Big]\, ,
\end{equation}
where the parameter $n$ is defined as follows,
\begin{equation}\label{nparameter}
n=\frac{m+2}{m+1}\, .
\end{equation}
One can easily calculate the first slow-roll index in this
minimally coupled scalar field framework, and we have,
\begin{equation}\label{firstslowrollfinal}
\epsilon_1=\frac{1}{1+(m+2) N}\, .
\end{equation}
Accordingly, the scalar spectral index is easily evaluated
\begin{equation}\label{scalarasfunctionofm}
n_{\mathcal{S}}=1-\frac{m+4}{1+(m+2) N}\, ,
\end{equation}
and the tensor-to-scalar ratio reads,
\begin{equation}\label{tensortoscalarratio}
r=\frac{16}{1+(m+2) N}\, .
\end{equation}
The inflationary phenomenology depends solely on one parameter,
namely $m$. At leading order in $N$, the observational indices
become,
\begin{equation}\label{nslargeN}
n_{\mathcal{S}}\simeq 1-\frac{m+4}{(m+2) N}+\frac{m+4}{(m+2)^2
N^2}\, ,
\end{equation}
\begin{equation}\label{rlargeN}
r\simeq \frac{16}{(m+2) N}-\frac{16}{(m+2)^2 N^2}\, .
\end{equation}
One can easily check that an ACT-compatible viable inflationary
phenomenology is obtained by choosing for example $m=6$ and
$N=55.5$, in which case we have $n_{\mathcal{S}}=0.977528$ and
$r=0.0359549$ which are both compatible with the ACT and the
BICEP/Planck data. The running of the spectral index can easily be
obtained by using Eq. (\ref{runningdefmainfinal}), so the
resulting expression for the running of the spectral index is,
\begin{equation}\label{runningfinalmodelanalytic}
a_s=\frac{m+4}{N \Big(1+(m+2) N\Big)}\, .
\end{equation}
Now apparently running of the spectral index for this model is
always positive, which is a unique feature for this model. And
also it is compatible with the ACT data, for example if
$(m,N)=(6,55.5)$ we get, $a_s=0.000404899$. Also for
asymptotically large values of the parameter $m$, the spectral
index saturates at $n_s\sim 1-\frac{1}{N}$ while $a_s\sim
\frac{1}{N^2}$ and the tensor-to-scalar ratio is almost zero, so
for sixty $e$-foldings we have $n_s\sim 0.98$ and $a_s\sim
0.000277778$. This is an example of a viable inflationary scalar
field theory with an always positive running of the spectral
index. This theory is unconventional though, since inflation is
singular, although if matter fields are added, the pressure
singularity is avoided even classically, or it is possible that
quantum effects might help avoid the pressure singularity if
classical matter fields are absent.

\section{Conclusions}

In this paper we investigated the possibility of having a positive
running of the spectral index in Einstein and Modified gravity. We
considered the mainstream theory of inflation in the context of
Einstein gravity, namely minimally coupled scalar field theory and
from modified gravities, we considered $F(R)$ gravity and
Einstein-Gauss-Bonnet gravity, which are both string motivated
theories. In the case of both minimally coupled scalar field
theory and $F(R)$ gravity, we demonstrated that it is quite
difficult to construct a viable inflationary theory which has a
positive running of the spectral index, and we showed the
objective problems one must deal with. For the scalar theory, some
unconventional scenarios might lead to a viable inflationary
theory with a positive running of the spectral index, and we
presented a singular inflation scenario, which can be solved
analytically, and which leads to a pressure singularity, if no
other matter fluids are present, except for the inflaton. In the
case of Einstein-Gauss-Bonnet theory, the situation is much
simpler and the positive running of the spectral index becomes a
model dependent issue, with no particular complications. We found
several Einstein-Gauss-Bonnet models that are compatible with the
GW170817 event and which lead to a viable inflationary era with
positive running of the spectral index and we presented a simple
one for brevity. We believe that in the case that the ACT data on
the running of the spectral index are confirmed, and the running
of the spectral index is actually positive, this will put in peril
many conventional inflationary models, but we demonstrated that
there exist simple models in terms of modified gravity, or nuanced
scalar field models that can actually produce naturally a positive
running of the spectral index. We aim to present several more
models of modified gravity with positive running of the spectral
index in a future work.

\end{document}